\documentclass{article}

\usepackage{graphicx}       
\usepackage{dcolumn}        
\usepackage{bm}             
\usepackage{subcaption}
\usepackage{hyperref}
\usepackage{amsmath}
\usepackage[symbol]{footmisc}
\usepackage[margin=1in]{geometry}

\usepackage[backend=biber,sorting=none,giveninits=true,sortcites=true]{biblatex}
\DeclareNameAlias{default}{family-given}    
\DeclareFieldFormat[article,incollection,inproceedings]{title}{#1}
\DeclareFieldFormat*{title}{#1}
\begin{document}

\title{Observation of Flat Bands in Type-II Weyl Semimetal TaRhTe\textsubscript{4}}
\author{Harry Rankin\textsuperscript{1,2}, Tyler J. Slade\textsuperscript{1,2}, Benjamin Schrunk\textsuperscript{1,2},\\ Yevhen Kushnirenko\textsuperscript{1,2,$\dag$}, Andrew Eaton\textsuperscript{1,2},\\ K. U. R. R. S. Rathnayaka\textsuperscript{1,2}, Maxwell Doyle\textsuperscript{1,2},\\ Lin-Lin Wang\textsuperscript{1,2}, Paul C. Canfield\textsuperscript{1,2}, Adam Kaminski\textsuperscript{1,2,*}}
\date{\today}

\maketitle

\begin{abstract}
Flat bands have been theoretically predicted for decades but have only recently been realized in quantum materials such as magic-angle twisted bilayer graphene, kagome and Lieb lattices, and rare-earth metal compounds. To date, only twisted layered materials have enabled tuning of flat-band energies near the electronic chemical potential, thereby influencing transport and thermodynamic properties. Here, we report the presence of flat bands near the chemical potential in bulk TaRhTe\textsubscript{4}, a noncentrosymmetric van-der Waals type-II Weyl semimetal. Flat bands are rarely observed in Weyl semimetals, particularly in nonmagnetic bulk systems, and the observed flat bands were not predicted by density functional theory calculations. TaRhTe\textsubscript{4} therefore provides a platform in which nontrivial topology coexists with flat bands near the Fermi level, as evidenced by our angle-resolved photoemission spectroscopy measurements.
\end{abstract}

\noindent
$^{1}$Ames National Laboratory, U.S. Department of Energy, Ames, Iowa 50011, USA \\ 
$^{2}$Department of Physics and Astronomy, Iowa State University, Ames, Iowa 50011, USA \\
$^*$\href{mailto:adamkam@ameslab.gov}{adamkam@ameslab.gov} \\
$\dag$ New address: Department of Physics, The Pennsylvania State University, University Park, PA, USA \\

\section*{Introduction}

Dispersionless electronic bands, or flat bands, have been predicted for decades \cite{Sutherland_1986, Wu_2007, Morell_2010}. They have recently been observed across a range of systems, driven by various mechanisms. For example, flat bands have been observed in magic-angle twisted bilayer graphene due to interlayer hybridization \cite{Cao_2018_1, Cao_2018_2, Lisi_2020, Bistritzer_2011}, kagome and Lieb lattices due to destructive wavefunction interference \cite{Ochi_2015, Nytko_2008, Kang_2020, Yang_2023}, and some rare-earth compounds because of hybridization between the $f$-level electrons and conduction electrons \cite{Ramankutty_2016, Yano_2008, Danzenbacher_2009}. Flat bands are highly unusual electronic states characterized by a range of momentum for which the energy is constant. These bands have a large density of states, and can lead to exotic electronic properties when located near the Fermi level, such as superconductivity, magnetism, and anomalous magnetotransport. Materials hosting these phenomena provide ideal platforms for studying systems where the independent-electron or weakly interacting-electron models fail \cite{Kang_2020, Si_2010, Tsui_1982, Cao_2018_1, Cao_2018_2}.

Another recent area of interest in quantum materials is the study of topological materials. Characterizing insulators, superconductors, and semimetals \cite{Kane_2010, Bansil_2016} by analyzing their band structure through the lens of topology has proven to be a powerful approach. Within the broad category of ``topologically nontrivial'' materials, topological semimetals are especially interesting because some types have electronic quasiparticles that satisfy relativistic wave equations for fermions, either the Dirac equation for Dirac semimetals \cite{Dirac_1928} or the Weyl equation for Weyl semimetals \cite{Weyl_1929}. The Weyl equation was initially intended to describe massless chiral fermions, which as free particles have not yet been observed in nature. However, these solid-state systems contain electronic quasiparticles with dispersion relations that these equations can describe \cite{Kane_2010, Bansil_2016}. In particular, Weyl semimetals contain points in reciprocal space with an inherent chirality where two nondegenerate linear bands touch \cite{Soluyanov_2015}. Flat bands have rarely been seen in nonmagnetic Weyl semimetals, especially in bulk crystals \cite{Xu_2020}.

The van der Waals compound TaIrTe\textsubscript{4} has received substantial attention recently, partially because two groups predicted it to realize a type-II Weyl semimetal state with only four Weyl points, the minimum possible for a noncentrosymmetric compound \cite{Haubold_2017, Belopolski_2017}. However, other groups have debated this \cite{Zhou_2018}. TaIrTe\textsubscript{4} exhibits many exotic phenomena, including the nonlinear Hall effect, persistent up to room temperature \cite{Kumar_2021}, and surface superconductivity below 1.54 K \cite{Xing_2019}. Whereas TaIrTe\textsubscript{4} has been extensively studied, the isostructural compound TaRhTe\textsubscript{4} remains underexplored; therefore, we have synthesized it and measured its electronic structure using angle-resolved photoemission spectroscopy (ARPES).

\begin{figure}[ht]
    \centering
    \includegraphics[width=\columnwidth]{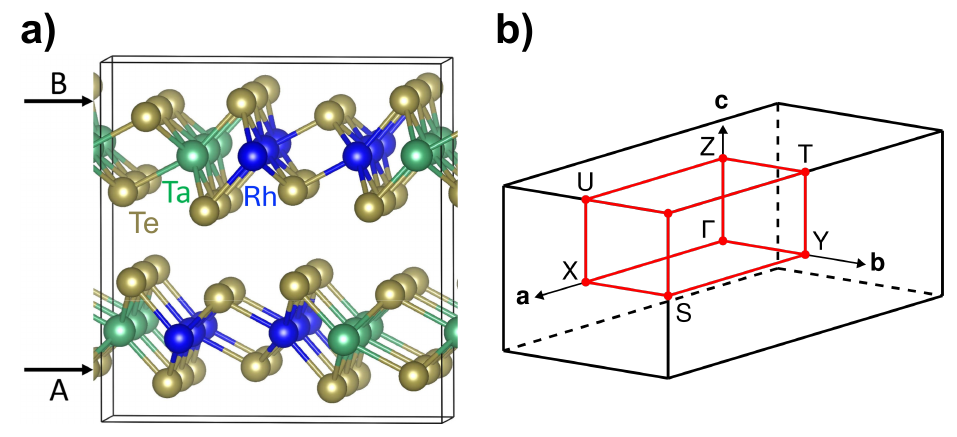}
    \caption{\textbf{The crystal structure and Brillouin zone of TaRhTe\textsubscript{4}.} \textbf{a} The crystal structure of TaRhTe\textsubscript{4}. Because it is noncentrosymmetric and layered, there are two different surface terminations, labeled by A and B here. \textbf{b} The Brillouin zone of TaRhTe\textsubscript{4}.}
    \label{fig:Crystal_Structure}
\end{figure}

Previous work on TaRhTe\textsubscript{4} includes density functional theory (DFT) studies of few-layer and bulk crystals, identification of the Weyl points, growth and crystal structure characterization, and investigation of magnetotransport properties compatible with the chiral anomaly \cite{Zhang_2024, Liu_2016, Shipunov_2021, Behnami_2025, Sadhukhan_2025}. Transport measurements on TaRhTe\textsubscript{4} show no indication of superconductivity down to 8 K \cite{Behnami_2025}, whereas TaIrTe\textsubscript{4} is reported to exhibit surface superconductivity with critical temperatures up to 1.54 K \cite{Xing_2019}.

Like TaIrTe\textsubscript{4}, the rhodium compound crystallizes in the noncentrosymmetric orthorhombic P$mn$2$_1$ structure (space group 31). The crystal structure is shown in Fig. \ref{fig:Crystal_Structure}a, and the Brillouin zone (BZ) in Fig. \ref{fig:Crystal_Structure}b. The crystals are quasi-two-dimensional with van der Waals bonds along the $c$ axis. The lattice constants have been characterized by two different groups: Shipunov et al. determined them to be $a=3.756$ \AA, $b=12.548$ \AA, and $c=13.166$ \AA \cite{Shipunov_2021}, and Mar et al. found values of $a=3.78$ \AA, $b=12.66$ \AA, and $c=13.19$ \AA \cite{Mar_1992}. The noncentrosymmetric and layered crystal structure gives rise to two different terminations, labeled A and B in Fig. \ref{fig:Crystal_Structure}a. The distinct terminations are also predicted in the isostructural TaIrTe\textsubscript{4} \cite{Koepernik_2016,Haubold_2017}.

\section*{Results}

\begin{figure}[ht]
    \centering
    \includegraphics[width=\columnwidth]{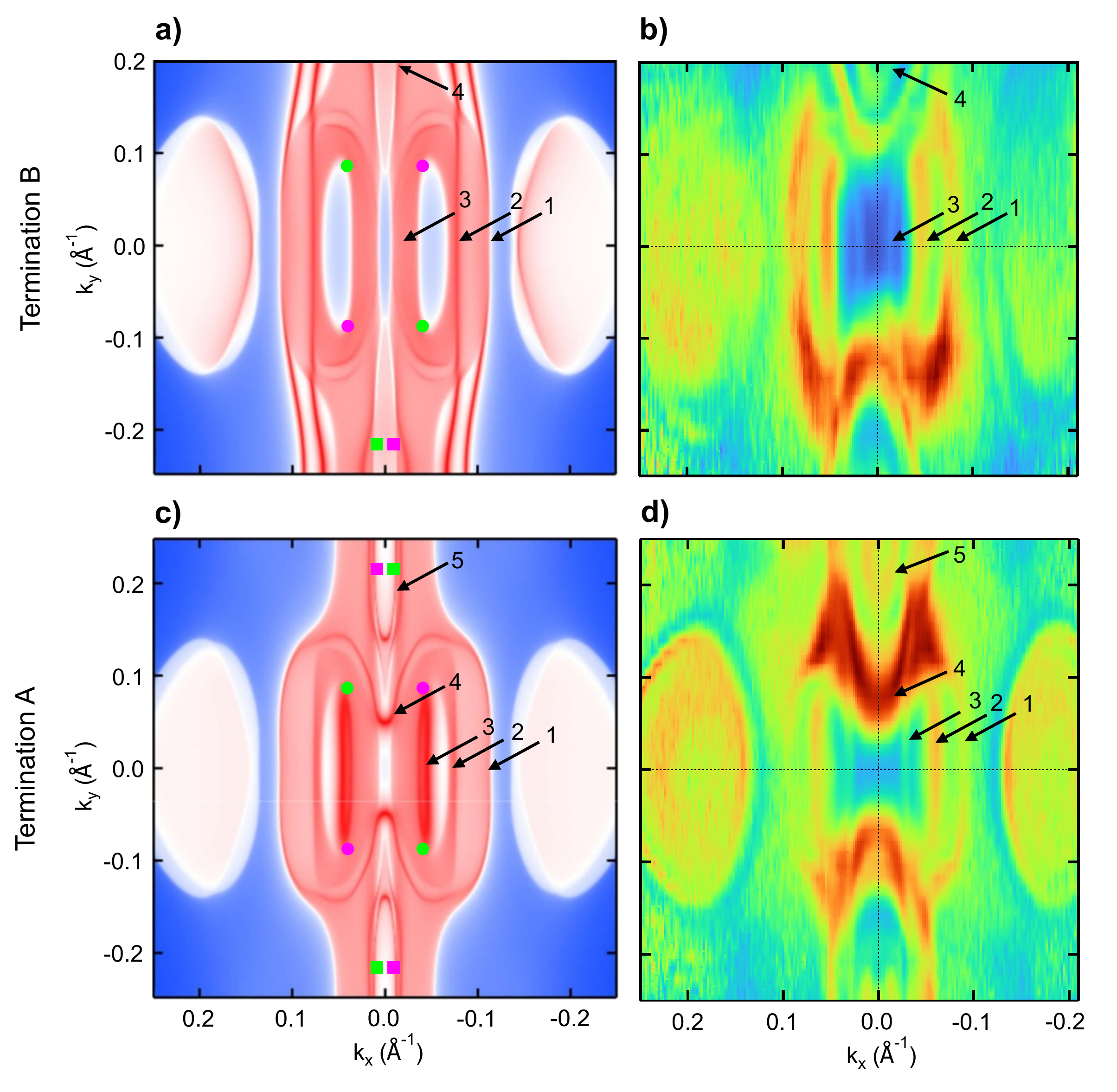}
    \caption{\textbf{Fermi surfaces of TaRhTe\textsubscript{4}.} \textbf{a,b} DFT-calculated and ARPES Fermi surfaces for termination B. The green and purple icons show the theoretical momentum positions of the Weyl points. The green WP projections have chirality -1 and the purple WP projections have chirality +1. The squares depict WP projections with predicted energy 13 meV above $E_F$, whereas the circles represent WP projections with energy 50 meV below $E_F$. \textbf{c,d} DFT-calculated and ARPES Fermi surfaces for termination A. Note that in this termination, we obtained data out to $k_y\approx 0.25$ 1/\AA, whereas for termination B, we only obtained data out to $k_y\approx 0.20$ 1/\AA.}
    \label{fig:FS}
\end{figure}
\begin{figure}[ht]
    \centering
    \includegraphics[width=\columnwidth]{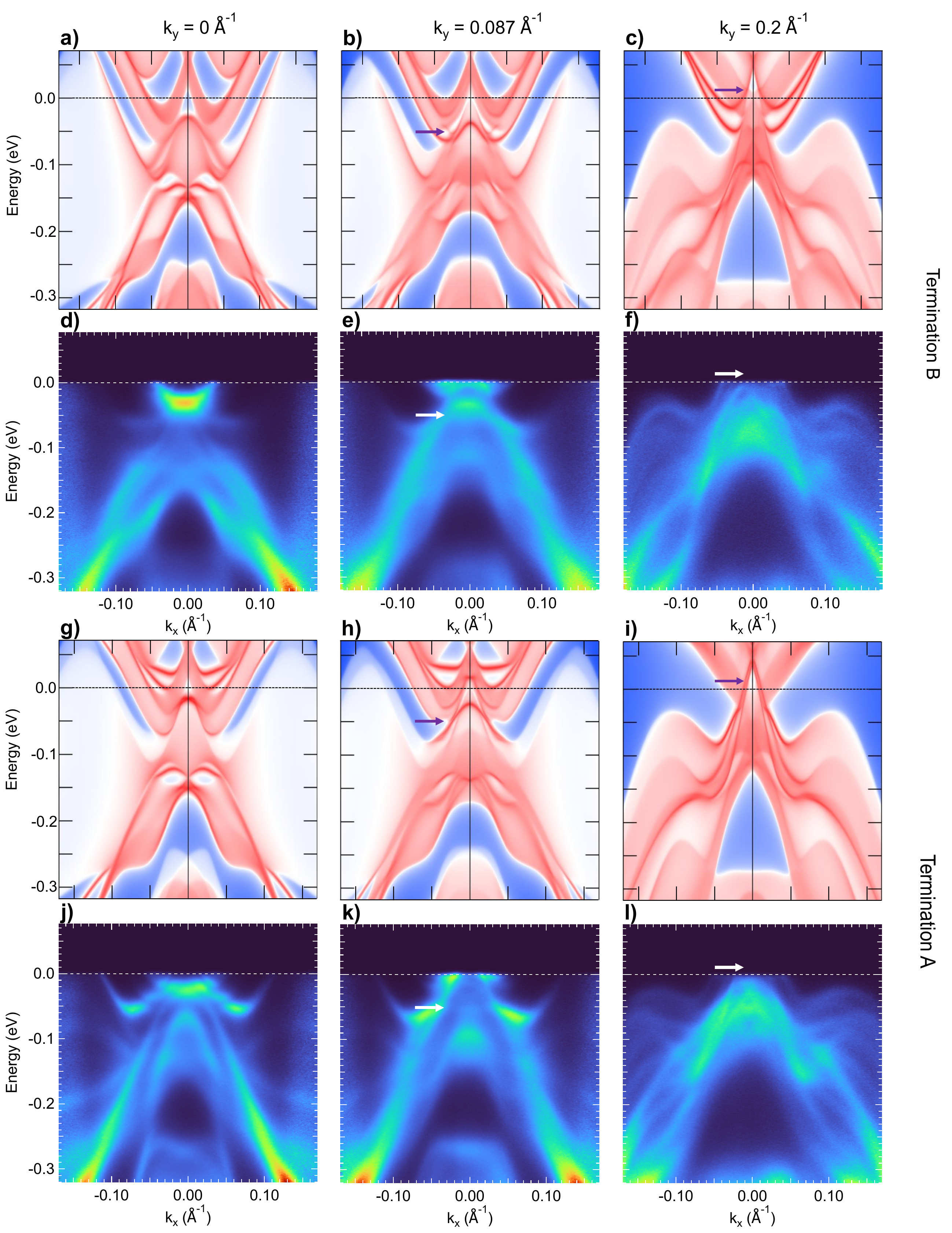}
    \caption{\textbf{DFT and ARPES comparison of several energy-momentum cuts near the WP locations on each termination.} \textbf{a-c} DFT-calculated energy-momentum cuts through termination B. The purple arrows indicate the theoretical locations of the WPs. \textbf{d-f} ARPES-measured energy-momentum cuts through termination B. The white arrows are placed at the same location as the purple arrows in the DFT calculations. \textbf{g-i} DFT-calculated energy-momentum cuts through termination B.
    \textbf{j-l} ARPES-measured energy-momentum cuts through termination B.}
    \label{fig:FS_cuts}
\end{figure}

The Fermi surfaces (FSs) of TaRhTe\textsubscript{4} predicted by DFT and measured by ARPES are shown in Fig. \ref{fig:FS}. DFT predicts two different Fermi surfaces corresponding to the two different surface terminations, shown in Fig. \ref{fig:Crystal_Structure}a. The ARPES data reveal two distinct electronic structures; contrast Fig. \ref{fig:FS_cuts} panels d-f with panels j-l. Careful comparison of the ARPES spectra and DFT calculations allowed assignment of each data set to the most likely corresponding calculated electronic structure, with the matched pairs shown in Fig. \ref{fig:FS}. For example, the surface states in termination B, marked by arrows 1 and 2 in Fig. \ref{fig:FS}a, stretch throughout the entire range of $k_y$. The ARPES spectra in Fig. \ref{fig:FS}b, especially the state marked by arrow 1, are more consistent with the termination B prediction in panel a. In the other termination, there are no similar states that stretch through all of $k_y$ in the manner predicted in panel a. Instead, we see that the FS narrows at the top and bottom, which is more compatible with what is predicted in Fig. \ref{fig:FS}c. Additionally, in termination A, the strong surface state labeled by arrow 4 appears quite clearly in both the DFT and ARPES data. While there are some features in the ARPES spectra that do not match the DFT predictions, and vice versa, these pairs as shown in Fig. \ref{fig:FS} provide the closest agreement between the two FSs.


\clearpage

The calculations predict eight Weyl points in TaRhTe\textsubscript{4}, the locations of which are contained in Table 1. Four WPs lie 13 meV above the electronic Fermi energy $E_F$, unresolvable through ordinary ARPES, and are labeled by the green and purple squares. Four WPs lie 50 meV below $E_F$; however, resolving these band crossings is challenging due to the overlap of several bands and the integration over a finite $k_z$ range inherent to the ARPES process.

\begin{table}
\centering
\caption{Location of Weyl points according to DFT calculations.}
\begin{tabular}{|c|c|c|c|}
    \hline
    \textbf{WP Set} & Momentum (1/\AA) & Energy (eV) \\
    \hline
    1 & $(\pm 0.041, \pm 0.087, 0)$ & -0.050 \\
    \hline
    2 & $(\pm 0.009, \pm 0.216, 0)$ & +0.013 \\
    \hline
\end{tabular}
\end{table}

The FSs measured using ARPES in Fig. \ref{fig:FS}, panels b and d, show many similar features. We note that the oval pockets near the sides of the plotted areas are present in both terminations, in good agreement with the DFT calculations. Termination A in Fig. \ref{fig:FS}d yields excellent agreement with the calculations in Fig. 2c. The previously mentioned ``D'' surface states are clearly visible in the center, indicated by arrow 3. Further along the $k_x$ direction, several bands marked by areas of high intensity appear. The inner one corresponds to the flat edges of the dark D-shaped sheets within the larger calculated FS sheets, labeled by arrow 2 in the DFT calculations and the ARPES data. The outer band, labeled by arrow 1, coincides with the closed surface state in the calculations and resembles a horizontal hourglass. Following this band around towards the inner part, strong, sharply-curved features are seen in both the theoretical and experimental FSs at $k_x=0$ and $k_y\approx\pm 0.05$ \AA$^{-1}$, labeled by arrow 4. Further along the $k_y$ direction, thin half-oval features (arrow 5) appear at the top and bottom of the BZ. In the calculations, these correspond to surface states that connect the WPs (green and purple squares) at $13$ meV above $E_F$.

Fermi arcs connecting some WP pairs can be seen on each termination in Fig. \ref{fig:FS}. The WPs denoted by the green and purple squares are only 13 meV above $E_F$, and have parts of the Fermi arcs still seen on the FS. The Fermi arcs are labeled with arrow 4 for termination B and arrow 5 on termination A. The WPs denoted by the green and purple circles are 50 meV below $E_F$ and have no visible Fermi arcs connecting them on the FS because their binding energy is further away from the $E_F$ than the other set of WPs.

Termination B shows less obvious agreement with the theory. The general shape is similar to termination A and the calculations, with oval hole pockets on the outside and features that stretch across $k_y$ and bulge near the middle. The intense, weakly curved FS sheets outside of the center, labeled with arrows 1 and 2 in Fig. 2a,b, can be identified with the surface bands running through the entire BZ in the calculations; except in the experimental data (Fig. 2b), it is unclear if they extend to the BZ edge because brighter states dominate in these regions. Near the top and bottom of the FS map, two distinct parabolic-like bands run through the BZ edge and are similar to the features at the same location in the calculations, near arrow 4. However, in the ARPES data, there are two distinct bands, each of which appears to be a continuous line. In the calculations, the features near this area extend through the BZ and lose intensity, rather than curving inward and connecting. These features are more similar to those found in termination A data in Fig. 2d, but are shifted up in $k_y$. Indeed, the data for termination B does not extend as far in $k_y$ and is not as sharp as the other termination data. This is exemplified by the oval pockets, which are less pronounced in the data of termination B.

The ARPES FSs of the two different surfaces share many similar features. Fig. \ref{fig:FS_cuts} shows several energy-momentum cuts of each FS, highlighting their noticeable differences below $E_F$. Also shown are the theoretical locations of the Weyl points and the corresponding energy-momentum cuts. There is good agreement between DFT and ARPES, particularly below $E_F-0.05$ eV. Near the Fermi energy, the spectra deviate from the calculated band structure, in some cases significantly. The most interesting difference is the appearance of flat bands in the ARPES spectra, which is the extra feature between arrows 4 and 5 in Fig. \ref{fig:FS}d and will be discussed in detail later.

\clearpage

Fig. \ref{fig:FS_cuts} panels a-f show some energy-$k_x$ slices through termination B at $k_y=$0, 0.087, and 0.2 \AA$^{-1}$. The last two $k_y$ values correspond to the theoretical momenta of the WPs, as listed in Table 1. The match between calculation and experiment is generally very good: most of the significant features of the data can be easily identified with the calculations. The finite resolution of ARPES sometimes leads to the appearance of single, strong bands that are actually due to the overlap of several closely spaced bands. This is evident in Fig. \ref{fig:FS_cuts}, panels a and d. The DFT plot in Fig. \ref{fig:FS_cuts}a shows multiple surface states near the bottom of the plot, between $k_x=0.10$ \AA$^{-1}$ and $k_x=0.15$ \AA$^{-1}$, but in the ARPES data in Fig. \ref{fig:FS_cuts}d, these appear to be a single strong band: the momentum distribution curves show a single peak.

\begin{figure}[ht]
    \centering
    \includegraphics[width=\columnwidth]{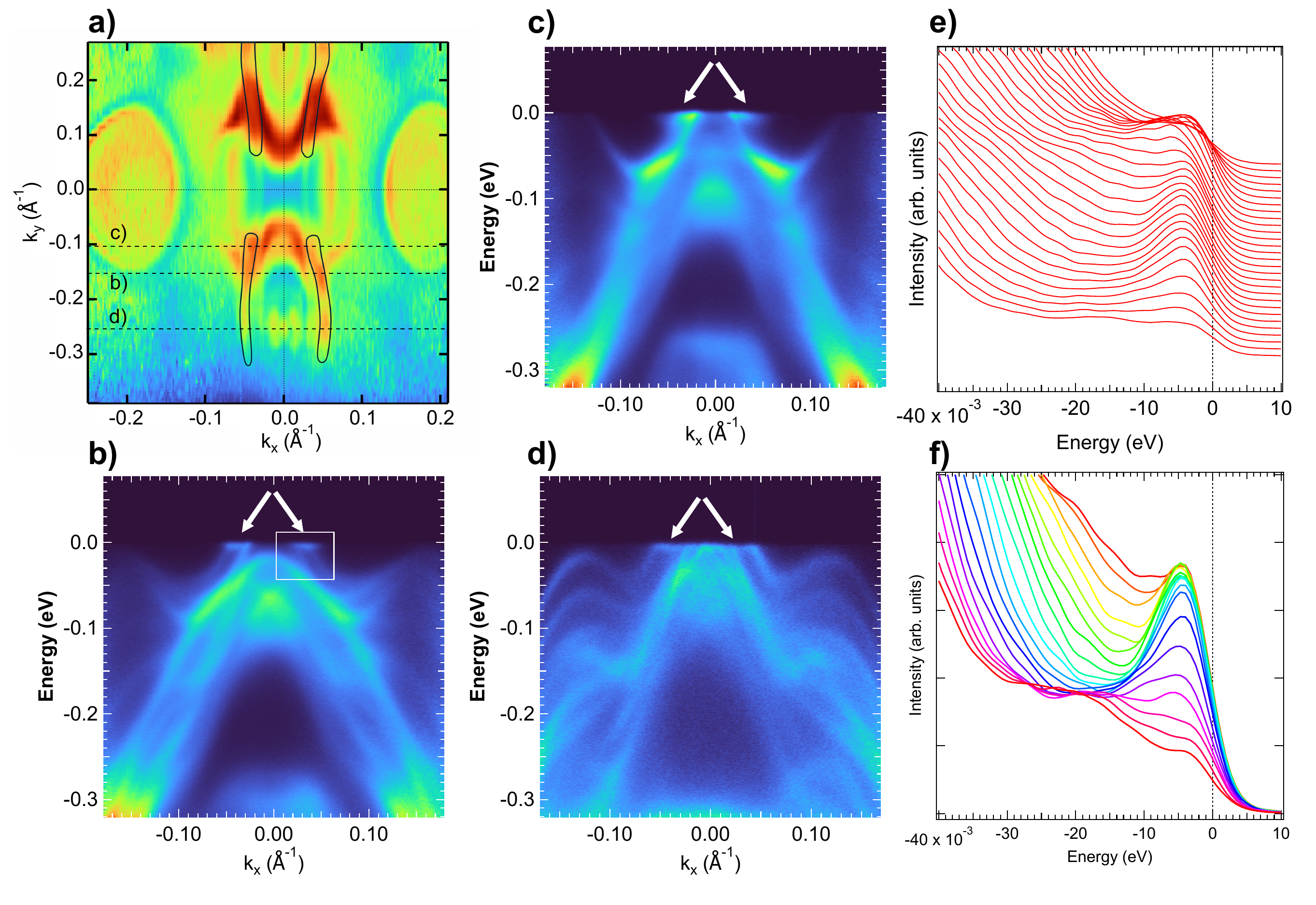}
    \caption{\textbf{Flat bands in the Fermi surface of termination A.} \textbf{a} FS of termination A, showing how the flat bands extend throughout most of the Brillouin zone. \textbf{b,c,d} Several energy-momentum cuts in which the flat bands are visible. The flat bands are shown by the white arrows. \textbf{e} EDCs of panel (\textbf{b}), focused on the area enclosed by the smaller white frame. \textbf{f} EDCs of panel (\textbf{b}) without offset. The individual EDCs are colored distinctly only to highlight that the peaks occur at the same energy for each EDC. The peak of the flat bands occurs approximately 4 meV below the Fermi energy.}
    \label{fig:Flat_bands}
\end{figure}

Fig. \ref{fig:FS_cuts} panels g-l show the corresponding cuts on termination A. Again, the results agree well with the calculation; the shapes and curvatures of most general features coincide. However, for this termination, some features appear that are not in the calculations. Flat bands located directly below $E_F$ in Fig. \ref{fig:Flat_bands} b,c,d are present throughout much of the BZ, despite the absence of similar features in the calculations.

\clearpage

Fig. \ref{fig:Flat_bands} shows the flat bands throughout the FS of termination A. The flat bands form a structure that can be seen on the FS in Fig. \ref{fig:Flat_bands}a with a thin, black outline. Fig. \ref{fig:Flat_bands} panels b, c, and d show several energy-momentum cuts where the flat bands are visible and highlighted with white arrows. Fig. \ref{fig:Flat_bands}e shows the offset energy distribution curves (EDCs) of panel b, showing the momentum independence of the peak energy. Fig. \ref{fig:Flat_bands}f shows the same EDCs without the offset and over a smaller energy range. A detailed analysis shows that the peak of the flat bands is roughly 4 meV below $E_F$ and is effectively independent of $k_x$.

\begin{figure}[ht]
    \centering
    \includegraphics[width=\columnwidth]{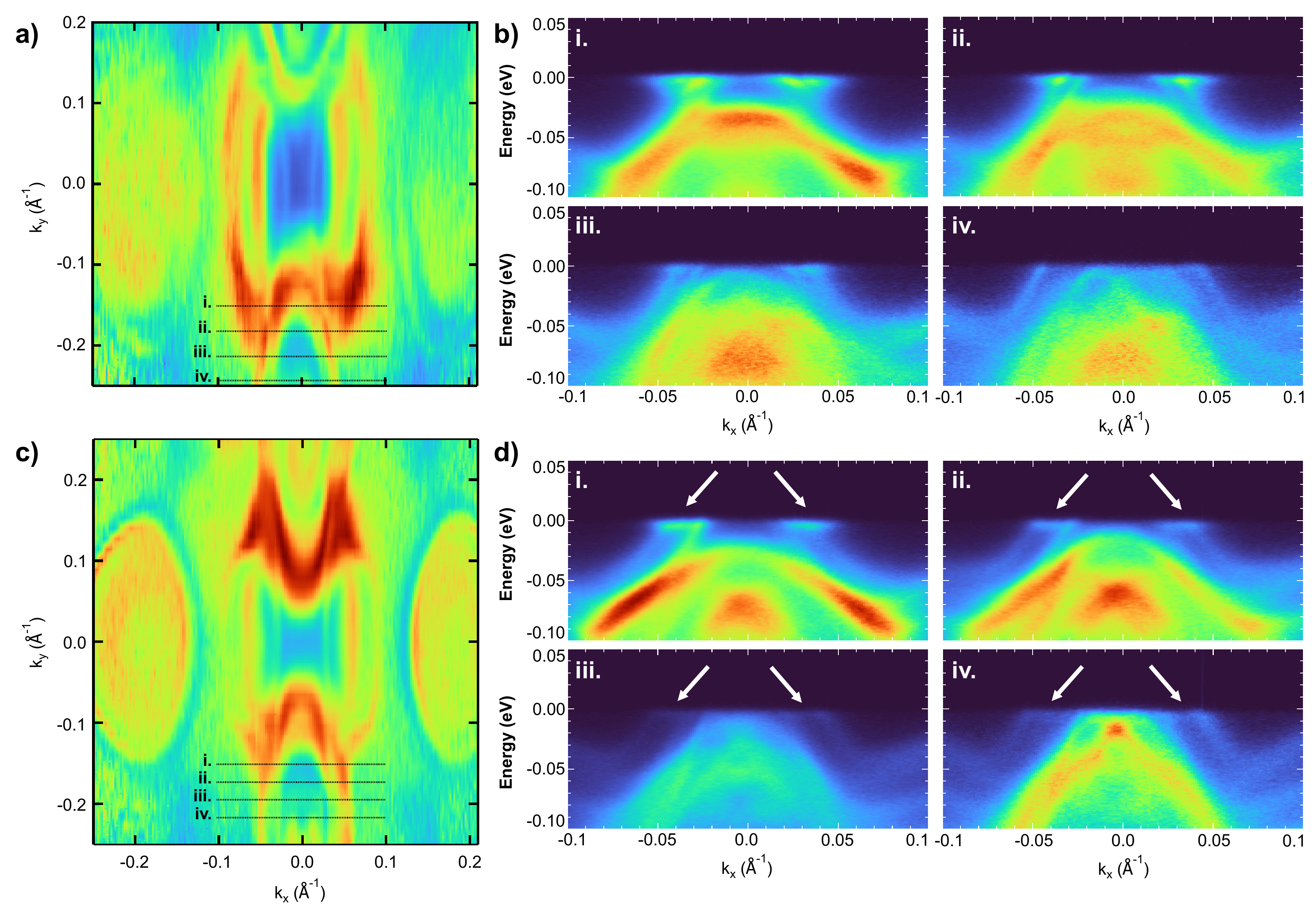}
    \caption{\textbf{Comparison of cuts in each termination near the location of the flat bands} \textbf{a} Termination B, where the lowercase roman numerals label the locations of the cuts shown in b i-iv. \textbf{b} Corresponding energy-momentum cuts. No flat bands are seen in termination B. \textbf{c} Termination A, which exhibits the flat bands. The lowercase roman numerals label the locations of the cuts shown in d i-iv. \textbf{d} Energy-momentum cuts at the locations labeled in panel c. The flat bands right below $E_F$ are indicated by the white arrows.}
    \label{fig:Termination_comparison_flat_bands}
\end{figure}

In Fig. \ref{fig:Termination_comparison_flat_bands}, we compare energy-momentum cuts at several $k_y$ which lie in the region of the flat bands in termination A. In panels b i-iii, there are notable features in the same location as the flat bands in the other termination; however, these features simply correspond to the bottom of an electron-like band. Indeed, in panel b iii, only the very bottom of the band can be seen, and in panel iv the band is above $E_F$. In Fig. \ref{fig:Termination_comparison_flat_bands}d i-iv, several momentum-energy cuts are shown in termination A, highlighting the flat bands.

\clearpage

\section*{Conclusion}
We have demonstrated that the Weyl semimetal TaRhTe\textsubscript{4} exhibits an unusual band structure that features both flat bands and Weyl points. The combination of these phenomena is rare and has not often been observed in nonmagnetic bulk crystals. The flat bands peak at 4 meV below $E_F$, allowing for chemical or electronic tuning to move them to the Fermi energy. The coexistence of flat-band physics and topological band structure makes this material a promising platform for investigating anomalous transport and emergent quantum phenomena.

\section*{Methods}

\subsection*{Crystal Growth}
Single crystals of TaRhTe\textsubscript{4} were grown using a solution/flux method \cite{Canfield_2020} with excess Te. Elemental Ta powder (Alfa-Aesar, 99.99\%), Rh powder (Alfa-Aesar, 99.99\%), and Te pieces (Alfa-Aesar, 99.99\%) were weighed in a 1:1:20 molar ratio of Ta:Rh:Te and placed into the growth end of a Canfield Crucible Set (CCS) \cite{Canfield_2016,LSP_Ceramics_2025}. The CCS was flame sealed in a fused silica ampoule. Prior to sealing, the ampoule was evacuated and then backfilled with ~1/6 atm Ar gas. The sealed ampoule was heated in a box furnace to 450 $^\circ$C over 5 h, then to 950 $^\circ$C in 10 h, held at this temperature for 10 h, and cooled to 600 $^\circ$C over 60 h. Upon reaching the final temperature, the ampoule was removed from the furnace, and the excess liquid phase was decanted using a centrifuge with a metal rotor and cups. After cooling to room temperature, the ampoules were opened to reveal thin, flake/blade-like crystals of TaRhTe\textsubscript{4}.

\subsection*{Band Structure Calculations}
We calculated the band structure of TaRhTe\textsubscript{4} in density functional theory \cite{Hohenberg_1964,Kohn_1965} (DFT) with spin-orbit coupling (SOC) in the PBE \cite{Perdew_1966} with modified Becke-Johnson \cite{Becke_2006} (mBJ) exchange-correlation functional at the experimental lattice parameters \cite{Mar_1992} using a plane-wave basis set and projector augmented wave method \cite{Blochl_1994}, as implemented in the Vienna Ab-initio Simulation Package (VASP) \cite{Kresse_1996a,Kresse_1996b}. In the DFT calculations, we used a kinetic energy cutoff of 229 eV, a $\Gamma$-centered Monkhorst-Pack \cite{Monkhorst_1976} (13×4×4) $k$-point mesh, and a Gaussian smearing of 0.05 eV. The experimental lattice constants \cite{Mar_1992} used are a=3.78 \AA, b=12.66 \AA, c=13.19 \AA. Using maximally localized Wannier functions \cite{Marzari_1997,Souza_2001}, tight-binding models were constructed to closely reproduce the band structure, including SOC within EF=$\pm$ 1eV with Ta s-d, Rh s-d, Te p orbitals. The surface spectral function and 2D Fermi surface were calculated with the surface Green’s function methods \cite{Sancho_1984,Sancho_1985} as implemented in WannierTools \cite{Wu_2018}.

\subsection*{ARPES}
Single-crystal samples of TaRhTe$_4$ were cleaved in situ under a vacuum of less than $5\times 10^{-11}$ Torr. Due to the van der Waals bonds along the (001) direction, the single crystals cleaved very easily, exposing an atomically flat surface. The light source used for measurements was a picosecond Ti:sapphire laser with a repetition rate of 80 MHz, and a fourth-harmonic generator was employed to achieve a final energy of 6.6 eV \cite{Jiang_2014}. The beam spot size was about 50 $\mu$m in diameter. Photoelectrons were collected and analyzed using a Scienta DA30 hemispherical electron analyzer. The energy and momentum resolutions were approximately 0.7 meV and 0.001 \AA$^{-1}$, respectively. The crystals were cleaved, and all data were collected at a temperature between 5 and 5.5 K.

\section*{Data Availability}
All raw data relevant to this work are available at the Harvard Dataverse data repository:\\  https://doi.org/10.7910/DVN/C0I1DR.

\printbibliography

@article{Weyl_1929,
    title={Elektron und gravitation. I},
    volume={56},
    DOI={10.1007/bf01339504},
    number={5–6},
    journal={Z. Phys.},
    author={Weyl, H.},
    year={1929},
    month={5},
    pages={330–352}
}

@article{Sutherland_1986,
    title = {Localization of electronic wave functions due to local topology},
    author = {Sutherland, B.},
    journal = {Phys. Rev. B},
    volume = {34},
    issue = {8},
    pages = {5208--5211},
    numpages = {0},
    year = {1986},
    month = {Oct},
    publisher = {American Physical Society},
    doi = {10.1103/PhysRevB.34.5208},
    %url = {https://link.aps.org/doi/10.1103/PhysRevB.34.5208}
}

@article{Wu_2007,
    title = {Flat Bands and Wigner Crystallization in the Honeycomb Optical Lattice},
    author = {Wu, C. and Bergman, D. and Balents, L. and Das Sarma, S.},
    journal = {Phys. Rev. Lett.},
    volume = {99},
    issue = {7},
    pages = {070401},
    numpages = {4},
    year = {2007},
    month = {Aug},
    publisher = {American Physical Society},
    doi = {10.1103/PhysRevLett.99.070401},
    %url = {https://link.aps.org/doi/10.1103/PhysRevLett.99.070401}
}

@article{Morell_2010,
    title = {Flat bands in slightly twisted bilayer graphene: Tight-binding calculations},
    author = {Su\'arez Morell, E. and Correa, J. D. and Vargas, P. and Pacheco, M. and Barticevic, Z.},
    journal = {Phys. Rev. B},
    volume = {82},
    issue = {12},
    pages = {121407},
    numpages = {4},
    year = {2010},
    month = {Sep},
    publisher = {American Physical Society},
    doi = {10.1103/PhysRevB.82.121407},
    %url = {https://link.aps.org/doi/10.1103/PhysRevB.82.121407}
}

@article{Kane_2010,
    title = {Colloquium: Topological insulators},
    author = {Hasan, M. Z. and Kane, C. L.},
    journal = {Rev. Mod. Phys.},
    volume = {82},
    issue = {4},
    pages = {3045--3067},
    numpages = {0},
    year = {2010},
    month = {Nov},
    publisher = {American Physical Society},
    doi = {10.1103/RevModPhys.82.3045},
    url = {https://link.aps.org/doi/10.1103/RevModPhys.82.3045}
}

@article{Bansil_2016,
    title = {Colloquium: Topological band theory},
    author = {Bansil, A. and Lin, H. and Das, T.},
    journal = {Rev. Mod. Phys.},
    volume = {88},
    issue = {2},
    pages = {021004},
    numpages = {37},
    year = {2016},
    month = {Jun},
    publisher = {American Physical Society},
    doi = {10.1103/RevModPhys.88.021004},
    %url = {https://link.aps.org/doi/10.1103/RevModPhys.88.021004}
}

@article{Dirac_1928,
    author = {Dirac, P. A. M.},
    title = {The quantum theory of the electron},
    journal = {Proc. R. Soc. Lond., Ser. A},
    volume = {117},
    number = {778},
    pages = {610-624},
    year = {1928},
    month = {02},
    issn = {0950-1207},
    doi = {10.1098/rspa.1928.0023},
    %url = {https://doi.org/10.1098/rspa.1928.0023},
    %eprint = {https://royalsocietypublishing.org/rspa/article-pdf/117/778/610/25048/rspa.1928.0023.pdf},
}

@article{Soluyanov_2015,
    title={Type-II Weyl semimetals},
    volume={527},
    DOI={10.1038/nature15768},
    number={7579},
    journal={Nat.},
    author={Soluyanov, A. A. and Gresch, D. and Wang, Z. and Wu, Q. and Troyer, M. and Dai, X. and Bernevig, B. A.},
    year={2015},
    month={11},
    pages={495–498}
}

@article{Haubold_2017,
    title = {Experimental realization of type-II Weyl state in noncentrosymmetric ${\mathrm{TaIrTe}}_{4}$},
    author = {Haubold, E. and Koepernik, K. and Efremov, D. and Khim, S. and Fedorov, A. and Kushnirenko, Y. and van den Brink, J. and Wurmehl, S. and B\"uchner, B. and Kim, T. K. and Hoesch, M. and Sumida, K. and Taguchi, K. and Yoshikawa, T. and Kimura, A. and Okuda, T. and Borisenko, S. V.},
    journal = {Phys. Rev. B},
    volume = {95},
    issue = {24},
    pages = {241108},
    numpages = {7},
    year = {2017},
    month = {6},
    publisher = {American Physical Society},
    doi = {10.1103/PhysRevB.95.241108},
    %url = {https://link.aps.org/doi/10.1103/PhysRevB.95.241108}
}

@article{Belopolski_2017,
    title = {Signatures of a time-reversal symmetric Weyl semimetal with only four Weyl points},
    author = {Belopolski, I. and Yu, P. and Sanchez, D. S. and Ishida, Y. and Chang, T.-R. and Zhang, S. S. and Xu, S.-Y. and Zheng, H. and Chang, G. and Bian, G. and Jeng, H.-T. and Kondo, T. and Lin, H. and Liu, Z. and Shin, S. and Hasan, M. Z.},
    journal = {Nat. Commun.},
    volume = {8},
    issue = {1},
    year = {2017},
    month = {10},
    publisher = {Nature Portfolio},
    doi = {10.1038/s41467-017-00938-1},
    %url = {https://www.nature.com/articles/s41467-017-00938-1}
}

@article{Kumar_2021,
    title={Room-temperature nonlinear Hall effect and wireless radiofrequency rectification in Weyl semimetal TaIrTe\textsubscript{4}},
    volume={16},
    DOI={10.1038/s41565-020-00839-3},
    number={4},
    journal={Nat. Nanotechnol.},
    author={Kumar, D. and Hsu, C.-H. and Sharma, R. and Chang, T.-R. and Yu, P. and Wang, J. and Eda, G. and Liang, G. and Yang, H.},
    year={2021},
    month={1},
    pages={421–425}
}

@article{Xing_2019,
    author = {Xing, Y. and Shao, Z. and Ge, J. and Luo, J. and Wang, J. and Zhu, Z. and Liu, J. and Wang, Y. and Zhao, Z. and Yan, J. and Mandrus, D. and Yan, B. and Liu, X.-J. and Pan, M. and Wang, J.},
    title = {Surface superconductivity in the type II Weyl semimetal TaIrTe\textsubscript{4}},
    journal = {Natl. Sci. Rev.},
    volume = {7},
    number = {3},
    pages = {579-587},
    year = {2019},
    month = {12},
    issn = {2095-5138},
    doi = {10.1093/nsr/nwz204},
    %url = {https://doi.org/10.1093/nsr/nwz204},
    %eprint = {https://academic.oup.com/nsr/article-pdf/7/3/579/38881219/nwz204.pdf},
}

@article{Zhou_2018,
    title={Coexistence of tunable Weyl points and topological nodal lines in ternary transition-metal telluride TaIrTe\textsubscript{4}},
    volume={97},
    DOI={10.1103/physrevb.97.241102},
    number={24},
    journal={Phys. Rev. B},
    author={Zhou, Xiaoqing and Liu, Qihang and Wu, QuanSheng and Nummy, Tom and Li, Haoxiang and Griffith, Justin and Parham, Stephen and Waugh, Justin and Emmanouilidou, Eve and Shen, Bing and Yazyev, Oleg V. and Ni, Ni and Dessau, Daniel},
    year={2018},
    month={6}
}

@article{Koepernik_2016,
  title = {${\mathrm{TaIrTe}}_{4}$: A ternary type-II Weyl semimetal},
  author = {Koepernik, K. and Kasinathan, D. and Efremov, D. V. and Khim, Seunghyun and Borisenko, Sergey and B\"uchner, Bernd and van den Brink, Jeroen},
  journal = {Phys. Rev. B},
  volume = {93},
  issue = {20},
  pages = {201101},
  numpages = {5},
  year = {2016},
  month = {5},
  publisher = {American Physical Society},
  doi = {10.1103/PhysRevB.93.201101},
  %url = {https://link.aps.org/doi/10.1103/PhysRevB.93.201101}
}

@article{Shipunov_2021,
    author = {Shipunov, G. and Piening, B. R. and Wuttke, C. and Romanova, T. A. and Sadakov, A. V. and Sobolevskiy, O. A. and Guzovsky, E. Yu. and Usoltsev, A. S. and Pudalov, V. M. and Efremov, D. V. and Subakti, S. and Wolf, D. and Lubk, A. and Büchner, B. and Aswartham, S.},
    title = {Layered van der Waals Topological Metals of TaTMTe\textsubscript{4} (TM = Ir, Rh, Ru) Family},
    journal = {J. Phys. Chem. Lett.},
    volume = {12},
    number = {28},
    pages = {6730-6735},
    year = {2021},
    doi = {10.1021/acs.jpclett.1c01648},
    note ={PMID: 34264086},
    %URL = {https://doi.org/10.1021/acs.jpclett.1c01648},
    %eprint = {https://doi.org/10.1021/acs.jpclett.1c01648}
}

@article{Cao_2018_1,
    title={Correlated insulator behaviour at half-filling in magic-angle graphene superlattices},
    volume={556},
    DOI={10.1038/nature26154},
    number={7699},
    journal={Nat.},
    author={Cao, Yuan and Fatemi, Valla and Demir, Ahmet and Fang, Shiang and Tomarken, Spencer L. and Luo, Jason Y. and Sanchez-Yamagishi, Javier D. and Watanabe, Kenji and Taniguchi, Takashi and Kaxiras, Efthimios and Ashoori, Ray C. and Jarillo-Herrero, Pablo},
    year={2018},
    month={3},
    pages={80–84}
}

@article{Cao_2018_2,
    title={Unconventional superconductivity in magic-angle graphene superlattices},
    volume={556},
    DOI={10.1038/nature26160},
    number={7699},
    journal={Nat.},
    author={Cao, Yuan and Fatemi, Valla and Fang, Shiang and Watanabe, Kenji and Taniguchi, Takashi and Kaxiras, Efthimios and Jarillo-Herrero, Pablo},
    year={2018},
    month={3},
    pages={43–50}
}

@article{Lisi_2020,
    title={Observation of flat bands in twisted bilayer graphene},
    volume={17},
    DOI={10.1038/s41567-020-01041-x},
    number={2},
    journal={Nat. Phys.},
    author={Lisi, Simone and Lu, Xiaobo and Benschop, Tjerk and de Jong, Tobias A. and Stepanov, Petr and Duran, Jose R. and Margot, Florian and Cucchi, Irène and Cappelli, Edoardo and Hunter, Andrew and Tamai, Anna and Kandyba, Viktor and Giampietri, Alessio and Barinov, Alexei and Jobst, Johannes and Stalman, Vincent and Leeuwenhoek, Maarten and Watanabe, Kenji and Taniguchi, Takashi and Rademaker, Louk and van der Molen, Sense Jan and Allan, Milan P. and Efetov, Dmitri K. and Baumberger, Felix},
    year={2020},
    month={9},
    pages={189–193}
}

@article{Bistritzer_2011,
    author = {Rafi Bistritzer and Allan H. MacDonald},
    title = {Moiré bands in twisted double-layer graphene},
    journal = {Proceedings of the National Academy of Sciences},
    volume = {108},
    number = {30},
    pages = {12233-12237},
    year = {2011},
    doi = {10.1073/pnas.1108174108}
}

@article{Xu_2020,
    title={Electronic correlations and flattened band in magnetic Weyl semimetal candidate Co\textsubscript{3}So\textsubscript{2}S\textsubscript{2}},
    volume={11},
    DOI={10.1038/s41467-020-17234-0},
    number={1},
    journal={Nat. Commun.},
    author={Xu, Yueshan and Zhao, Jianzhou and Yi, Changjiang and Wang, Qi and Yin, Qiangwei and Wang, Yilin and Hu, Xiaolei and Wang, Luyang and Liu, Enke and Xu, Gang and Lu, Ling and Soluyanov, Alexey A. and Lei, Hechang and Shi, Youguo and Luo, Jianlin and Chen, Zhi-Guo},
    year={2020},
    month={8}
}

@article{Si_2010,
    title = {Heavy Fermions and Quantum Phase Transitions},
    author = {Si, Q. and Steglich, F.},
    journal = {Science},
    volume = {329},
    number = {5996},
    pages = {1161-1166},
    year = {2010},
    doi = {10.1126/science.1191195},
    %URL = {https://www.science.org/doi/abs/10.1126/science.1191195},
    %eprint = {https://www.science.org/doi/pdf/10.1126/science.1191195}
}

@article{Tsui_1982,
  title = {Two-Dimensional Magnetotransport in the Extreme Quantum Limit},
  author = {Tsui, D. C. and Stormer, H. L. and Gossard, A. C.},
  journal = {Phys. Rev. Lett.},
  volume = {48},
  issue = {22},
  pages = {1559--1562},
  numpages = {0},
  year = {1982},
  month = {5},
  publisher = {American Physical Society},
  doi = {10.1103/PhysRevLett.48.1559},
  %url = {https://link.aps.org/doi/10.1103/PhysRevLett.48.1559}
}

@article{Zhang_2024,
  title = {Layer dependent topological phases and transitions in ${\mathrm{TaRhTe}}_{4}$: From monolayer and bilayer to bulk},
  author = {Zhang, X. and Mao, N. and Janson, O. and van den Brink, J. and Ray, R.},
  journal = {Phys. Rev. Mater.},
  volume = {8},
  issue = {9},
  pages = {094201},
  numpages = {11},
  year = {2024},
  month = {9},
  publisher = {American Physical Society},
  doi = {10.1103/PhysRevMaterials.8.094201},
  %url = {https://link.aps.org/doi/10.1103/PhysRevMaterials.8.094201}
}

@article{Liu_2016,
    title={van der Waals Stacking-Induced Topological Phase Transition in Layered Ternary Transition Metal Chalcogenides},
    volume={17},
    DOI={10.1021/acs.nanolett.6b04487},
    number={1},
    journal={Nano Lett.},
    author={Liu, J. and Wang, H. and Fang, C. and Fu, L. and Qian, X.},
    year={2016},
    month={12},
    pages={467–475}
}

@article{Sadhukhan_2025,
    title={The orbital-driven topological phase transition and planar Hall responses in ternary tellurides Weyl semi-metals}, 
    author={Sadhukhan, B and Nag, T},
    year={2025},
    eprint={2509.19818},
    archivePrefix={arXiv},
    primaryClass={cond-mat.mtrl-sci},
    %url={https://arxiv.org/abs/2509.19818}
}

@article{Behnami_2025,
    title = {Signature of chiral anomaly in the Weyl semimetal ${\text{TaRhTe}}_{4}$},
    author = {Behnami, M. and Efremov, D. V. and Aswartham, S. and Shipunov, G. and Piening, B. R. and Blum, C. G. F. and Kocsis, V. and Dufouleur, J. and Pallecchi, I. and Putti, M. and B\"uchner, B. and Reichlova, H. and Caglieris, F.},
    journal = {Phys. Rev. B},
    volume = {112},
    issue = {4},
    pages = {045101},
    numpages = {7},
    year = {2025},
    month = {7},
    publisher = {American Physical Society},
    doi = {10.1103/n7lw-7xqj},
    %url = {https://link.aps.org/doi/10.1103/n7lw-7xqj}
}

@article{Jiang_2014,
    author = {Jiang, Rui and Mou, Daixiang and Wu, Yun and Huang, Lunan and McMillen, Colin D. and Kolis, Joseph and Giesber, Henry G., III and Egan, John J. and Kaminski, Adam},
    title = {Tunable vacuum ultraviolet laser based spectrometer for angle resolved photoemission spectroscopy},
    journal = {Rev. Sci. Instrum.},
    volume = {85},
    number = {3},
    pages = {033902},
    year = {2014},
    month = {03},
    issn = {0034-6748},
    doi = {10.1063/1.4867517},
    %url = {https://doi.org/10.1063/1.4867517},
    %eprint = {https://pubs.aip.org/aip/rsi/article-pdf/doi/10.1063/1.4867517/15603111/033902_1_online.pdf},
}

@article{Canfield_2016,
    author = {Canfield, P. C. and Kong, T. and Kaluarachchi, U. S. and Jo, N. H.},
    title = {Use of frit-disc crucibles for routine and exploratory solution growth of single crystalline samples},
    journal = {Philos. Mag.},
    volume = {96},
    number = {1},
    pages = {84--92},
    year = {2016},
    publisher = {Taylor \& Francis},
    doi = {10.1080/14786435.2015.1122248},
    %URL = {https://doi.org/10.1080/14786435.2015.1122248},
    %eprint = {https://doi.org/10.1080/14786435.2015.1122248}
}

@misc{LSP_Ceramics_2025,
    title={LSP Ceramics. Canfield Crucible Sets.},
    url={https://www.lspceramics.com/},
    note={Accessed September 2025}
}

@article{Hohenberg_1964,
    title = {Inhomogeneous Electron Gas},
    author = {Hohenberg, P. and Kohn, W.},
    journal = {Phys. Rev.},
    volume = {136},
    issue = {3B},
    pages = {B864--B871},
    numpages = {0},
    year = {1964},
    month = {11},
    publisher = {American Physical Society},
    doi = {10.1103/PhysRev.136.B864},
    %url = {https://link.aps.org/doi/10.1103/PhysRev.136.B864}
}

@article{Kohn_1965,
    title = {Self-Consistent Equations Including Exchange and Correlation Effects},
    author = {Kohn, W. and Sham, L. J.},
    journal = {Phys. Rev.},
    volume = {140},
    issue = {4A},
    pages = {A1133--A1138},
    numpages = {0},
    year = {1965},
    month = {11},
    publisher = {American Physical Society},
    doi = {10.1103/PhysRev.140.A1133},
    %url = {https://link.aps.org/doi/10.1103/PhysRev.140.A1133}
}

@article{Perdew_1966,
    title = {Generalized Gradient Approximation Made Simple},
    author = {Perdew, J. P. and Burke, K. and Ernzerhof, M.},
    journal = {Phys. Rev. Lett.},
    volume = {77},
    issue = {18},
    pages = {3865--3868},
    numpages = {0},
    year = {1996},
    month = {10},
    publisher = {American Physical Society},
    doi = {10.1103/PhysRevLett.77.3865},
    %url = {https://link.aps.org/doi/10.1103/PhysRevLett.77.3865}
}

@article{Becke_2006,
    author = {Becke, A. D. and Johnson, E. R.},
    title = {A simple effective potential for exchange},
    journal = {J. Chem. Phys.},
    volume = {124},
    number = {22},
    pages = {221101},
    year = {2006},
    month = {06},
    issn = {0021-9606},
    doi = {10.1063/1.2213970},
    %url = {https://doi.org/10.1063/1.2213970},
    %eprint = {https://pubs.aip.org/aip/jcp/article-pdf/doi/10.1063/1.2213970/15385734/221101_1_online.pdf},
}

@article{Mar_1992,
    title = {Metal-metal vs tellurium-tellurium bonding in WTe2 and its ternary variants TaIrTe4 and NbIrTe4},
    author = {Mar, A. and Jobic, S. and Ibers, J. A.},
    journal = {J. Am. Chem. Soc.},
    volume = {114},
    number = {23},
    pages = {8963-8971},
    year = {1992},
    doi = {10.1021/ja00049a029},
    %URL = {https://doi.org/10.1021/ja00049a029},
    %eprint = {https://doi.org/10.1021/ja00049a029}
}

@article{Blochl_1994,
    title = {Projector augmented-wave method},
    author = {Bl\"ochl, P. E.},
    journal = {Phys. Rev. B},
    volume = {50},
    issue = {24},
    pages = {17953--17979},
    numpages = {0},
    year = {1994},
    month = {12},
    publisher = {American Physical Society},
    doi = {10.1103/PhysRevB.50.17953},
    %url = {https://link.aps.org/doi/10.1103/PhysRevB.50.17953}
}

@article{Kresse_1996a,
    title = {Efficient iterative schemes for \textit{ab initio} total-energy calculations using a plane-wave basis set},
    author = {Kresse, G. and Furthm\"uller, J.},
    journal = {Phys. Rev. B},
    volume = {54},
    issue = {16},
    pages = {11169--11186},
    numpages = {0},
    year = {1996},
    month = {10},
    publisher = {American Physical Society},
    doi = {10.1103/PhysRevB.54.11169},
    %url = {https://link.aps.org/doi/10.1103/PhysRevB.54.11169}
}

@article{Kresse_1996b,
    title = {Efficiency of ab-initio total energy calculations for metals and semiconductors using a plane-wave basis set},
    author = {Kresse, G. and Furthm\"uller, J.},
    journal = {Comput. Mater. Sci.},
    volume = {6},
    number = {1},
    pages = {15-50},
    year = {1996},
    issn = {0927-0256},
    doi = {https://doi.org/10.1016/0927-0256(96)00008-0},
    %url = {https://www.sciencedirect.com/science/article/pii/0927025696000080}
}

@article{Monkhorst_1976,
    title = {Special points for Brillouin-zone integrations},
    author = {Monkhorst, H. J. and Pack, J. D.},
    journal = {Phys. Rev. B},
    volume = {13},
    issue = {12},
    pages = {5188--5192},
    numpages = {0},
    year = {1976},
    month = {6},
    publisher = {American Physical Society},
    doi = {10.1103/PhysRevB.13.5188},
    %url = {https://link.aps.org/doi/10.1103/PhysRevB.13.5188}
}

@article{Marzari_1997,
    title = {Maximally localized generalized Wannier functions for composite energy bands},
    author = {Marzari, N. and Vanderbilt, D.},
    journal = {Phys. Rev. B},
    volume = {56},
    issue = {20},
    pages = {12847--12865},
    numpages = {0},
    year = {1997},
    month = {11},
    publisher = {American Physical Society},
    doi = {10.1103/PhysRevB.56.12847},
    %url = {https://link.aps.org/doi/10.1103/PhysRevB.56.12847}
}

@article{Souza_2001,
    title = {Maximally localized Wannier functions for entangled energy bands},
    author = {Souza, I. and Marzari, N. and Vanderbilt, D.},
    journal = {Phys. Rev. B},
    volume = {65},
    issue = {3},
    pages = {035109},
    numpages = {13},
    year = {2001},
    month = {12},
    publisher = {American Physical Society},
    doi = {10.1103/PhysRevB.65.035109},
    %url = {https://link.aps.org/doi/10.1103/PhysRevB.65.035109}
}

@article{Sancho_1984,
    title={Quick iterative scheme for the calculation of transfer matrices: application to Mo (100)},
    volume={14},
    ISSN={0305-4608},
    DOI={10.1088/0305-4608/14/5/016},
    number={5},
    journal={J. Phys. F: Met. Phys.},
    author={Lopez Sancho, M. P. and Lopez Sancho, J. M. and Rubio, J.},
    year={1984},
    pages={1205-1215}
}

@article{Sancho_1985,
    author = {Lopez Sancho, M. P. and Lopez Sancho, J. M. and Sancho, J. M. L. and Rubio, J.},
    title = {Highly convergent schemes for the calculation of bulk and surface Green functions},
    journal = {J. Phys. F: Met. Phys.},
    year = 1985,
    month = {4},
    volume = {15},
    number = {4},
    pages = {851-858},
    doi = {10.1088/0305-4608/15/4/009},
    %adsurl = {https://ui.adsabs.harvard.edu/abs/1985JPhF...15..851L}
}

@article{Wu_2018,
    title = {WannierTools: An open-source software package for novel topological materials},
    author = {Wu, Q and Zhang, S. and Song, H.-F. and Troyer, M. and Soluyanov, A. A.},
    journal = {Comput. Phys. Commun.},
    volume = {224},
    pages = {405-416},
    year = {2018},
    issn = {0010-4655},
    doi = {https://doi.org/10.1016/j.cpc.2017.09.033},
    %url = {https://www.sciencedirect.com/science/article/pii/S0010465517303442}
}

@article{Canfield_2020,
    doi = {10.1088/1361-6633/ab514b},
    url = {https://doi.org/10.1088/1361-6633/ab514b},
    year = {2019},
    month = {nov},
    publisher = {IOP Publishing},
    volume = {83},
    number = {1},
    pages = {016501},
    author = {Canfield, Paul C},
    title = {New materials physics},
    journal = {Reports on Progress in Physics}
}

@article{Ramankutty_2016,
    title = {Comparative study of rare earth hexaborides using high resolution angle-resolved photoemission},
    author = {S.V. Ramankutty and N. {de Jong} and Y.K. Huang and B. Zwartsenberg and F. Massee and T.V. Bay and M.S. Golden and E. Frantzeskakis},
    journal = {Journal of Electron Spectroscopy and Related Phenomena},
    volume = {208},
    pages = {43-50},
    year = {2016},
    note = {Special Issue: Electronic structure and function from state-of-the-art spectroscopy and theory},
    issn = {0368-2048},
    doi = {https://doi.org/10.1016/j.elspec.2015.11.009},
    url = {https://www.sciencedirect.com/science/article/pii/S0368204815002832}
}

@article{Yano_2008,
    title = {Electronic structure of $\mathrm{Ce}{\mathrm{Ru}}_{2}{X}_{2}$ $(X=\mathrm{Si},\mathrm{Ge})$ in the paramagnetic phase studied by soft x-ray ARPES and hard x-ray photoelectron spectroscopy},
    author = {Yano, M. and Sekiyama, A. and Fujiwara, H. and Amano, Y. and Imada, S. and Muro, T. and Yabashi, M. and Tamasaku, K. and Higashiya, A. and Ishikawa, T. and \ifmmode \bar{O}\else \={O}\fi{}nuki, Y. and Suga, S.},
    journal = {Phys. Rev. B},
    volume = {77},
    issue = {3},
    pages = {035118},
    numpages = {8},
    year = {2008},
    month = {Jan},
    publisher = {American Physical Society},
    doi = {10.1103/PhysRevB.77.035118},
    url = {https://link.aps.org/doi/10.1103/PhysRevB.77.035118}
}

@article{Danzenbacher_2009,
    title = {Hybridization Phenomena in Nearly-Half-Filled $f$-Shell Electron Systems: Photoemission Study of ${\mathrm{EuNi}}_{2}{\mathrm{P}}_{2}$},
    author = {Danzenb\"acher, S. and Vyalikh, D. V. and Kucherenko, Yu. and Kade, A. and Laubschat, C. and Caroca-Canales, N. and Krellner, C. and Geibel, C. and Fedorov, A. V. and Dessau, D. S. and Follath, R. and Eberhardt, W. and Molodtsov, S. L.},
    journal = {Phys. Rev. Lett.},
    volume = {102},
    issue = {2},
    pages = {026403},
    numpages = {4},
    year = {2009},
    month = {Jan},
    publisher = {American Physical Society},
    doi = {10.1103/PhysRevLett.102.026403},
    url = {https://link.aps.org/doi/10.1103/PhysRevLett.102.026403}
}

@article{Kang_2020,
    title={Topological flat bands in frustrated Kagome lattice CoSn},
    volume={11},
    DOI={10.1038/s41467-020-17465-1},
    number={1},
    journal={Nature Communications},
    author={Kang, Mingu and Fang, Shiang and Ye, Linda and Po, Hoi Chun and Denlinger, Jonathan and Jozwiak, Chris and Bostwick, Aaron and Rotenberg, Eli and Kaxiras, Efthimios and Checkelsky, Joseph G. and et al.},
    year={2020},
    month={Aug}
}

@article{Ochi_2015,
    title = {Robust flat bands in $R{\mathrm{Co}}_{5}$ ($R\phantom{\rule{0.16em}{0ex}}=\phantom{\rule{0.16em}{0ex}}\mathrm{rare}$ earth) compounds},
    author = {Ochi, Masayuki and Arita, Ryotaro and Matsumoto, Munehisa and Kino, Hiori and Miyake, Takashi},
    journal = {Phys. Rev. B},
    volume = {91},
    issue = {16},
    pages = {165137},
    numpages = {5},
    year = {2015},
    month = {Apr},
    publisher = {American Physical Society},
    doi = {10.1103/PhysRevB.91.165137}
}

@article{Nytko_2008,
    author = {Nytko, Emily A. and Helton, Joel S. and M\"uller, Peter and Nocera, Daniel G.},
    title = {A Structurally Perfect S = 1/2 Metal-Organic Hybrid Kagomé Antiferromagnet},
    journal = {Journal of the American Chemical Society},
    volume = {130},
    number = {10},
    pages = {2922-2923},
    year = {2008},
    doi = {10.1021/ja709991u},
    note ={PMID: 18275194}
}

\section*{Acknowledgements}
This work was supported by the U.S. Department of Energy, Office of Basic Energy Sciences, Division of Materials Science and Engineering. Ames National Laboratory is operated for the U.S. Department of Energy by Iowa State University under Contract No. DE-AC02-07CH11358.

\section*{Author Contributions}
T.J.S. and P.C.C. grew the samples. L.-L.W. performed DFT calculations. H.R., B.S., Y.K., A.E., K.U.R.R.S.R., M.D., and A.K. performed ARPES measurements and provided experimental support. The manuscript was drafted by H.R., B.S., Y.K., A.E., K.U.R.R.S.R., M.D., T.J.S., P.C.C., L.-L.W., and A.K. All authors discussed and commented on the manuscript.

\section*{Author Correspondents}
Tyler Slade (slade@iastate.edu) for materials information and requests and Adam Kaminski\\ (adamkam@ameslab.gov) for general inquiries.

\section*{Competing Interests}
The authors declare no competing interests.

\end{document}